\documentclass[12pt,a4paper]{article}
\usepackage{amssymb}
\usepackage[T2A]{fontenc}
\usepackage[cp866]{inputenc}
\usepackage[russian,english]{babel}
\usepackage[dvips,unicode]{hyperref}
\usepackage[dvips,unicode]{hyperref}
\begin{document}
\title{Inertial frames of reference, \\
space and time measurements, \\
and physical principles of special relativity revisited} 

\author{Andrew E. Chubykalo${}^\dagger$, Augusto Espinoza${}^\dagger$, and
B. P. Kosyakov${}^\ddagger$}

\maketitle

\begin{center}
{\it 
${}^\dagger$ Escuela de F${\acute\imath}$sica, Universidad
Aut\'onoma de Zacatecas, Apartado Postal C-580 Zacatecas 98068, ZAC, Mexico%
\\[0pt]
${}^\ddagger$ Russian Federal Nuclear Center, Sarov, 607190 Nizhnii Novgorod
Region, Russia \\[0pt]
\textrm{E-mail:} $\mathrm{achubykalo@yahoo.com.mx}$, 
$\mathrm{drespinozag@yahoo.com.mx}$, $\mathrm{kosyakov@vniief.ru}$} 
 
\end{center}

\def\emline#1#2#3#4#5#6{%
         \put(#1,#2){\special{em:moveto}}%
         \put(#4,#5){\special{em:lineto}}}
\def\newpic#1{}

\begin{abstract}
\noindent We give a critical analysis of the conceptual foundations of
special relativity. We formulate a simple operational criterion for
distinguishing between noninertial and inertial frames which is introduced
prior to geometry. We associate the concept of maximal velocity with the
existence of an upper bound for a set of rates of movers which travel in the
same direction. We define the standard scale for reading the time flow. We
refine the treatment of both Einstein's postulates, the principle of
relativity and constancy of the velocity of light. The proposed
``reconstruction'' of the geometry of Minkowski space will hopefully be
useful for the ongoing examination of possible Lorentz violations.
\end{abstract}


\section{Introduction}
\label{introduction} 
As is clear from the title, we address the issues
that are certain to be well-known to every physicist. ``Now what is to
discuss?'', the reader may wonder. The properties of space and time as well
as the procedure of their measurement are a matter for scientific inquiry at
all times. The information content increases and the knowledge structure is
rearranged. That is why, from time to time, one should revert to the formal
and conceptual aspects of the geometric essentials of our world if only to
ascertain that the edifice still rests on a solid base.

However, there is also a particular reason for an effort to be made to
re-examine the current status of geometric notions behind special
relativity, a hundred years after the advent of this theory. The idea of a 
\textit{failure of Lorentz invariance} continued to grow in popularity over
the last two decades. The appeal to this idea is due to some results in
string theory (for a review of these developments see \cite{Mattingly} and 
\cite{Jacobson}). In particular, Nathan Seiberg and Edward Witten 
\cite{Seiberg} were able to show that, in some low-energy limit, string theory
becomes a noncommutative field theory\footnote{Different approaches to 
constructing noncommutative field theories is
reviewed in \cite{Douglas}.} in which spacetime coordinates $x^{\mu}$ are
turned to operators satisfying the commutation relation 
\begin{equation}
[x^\mu, x^\nu] = i\theta^{\mu\nu}.  \label{noncomm}
\end{equation}
Here, $\theta^{\mu\nu}$ is a constant antisymmetric tensor which is
responsible for the departure from the ordinary picture with commuting
coordinates. In such theories with noncommuting coordinates, Lorentz
invariance is violated. This is most easily understood if one considers the
vacuum expectation value of Eq.~(\ref{noncomm}) which shows a manifest
Lorentz violation\footnote{The vacuum $|0\rangle$ is, by definition 
(see, e.~g., \cite{Wightman}), the
unique Lorentz invariant normalized state: $\exp(i\,\omega^{\alpha\beta}M_{%
\alpha\beta})\,|0\rangle = |0\rangle$, $\langle 0|0\rangle=1$, where $%
\omega^{\alpha\beta}$ and $M_{\alpha\beta}$ are, respectively, parameters
and generators of the Lorentz group. The vacuum expectation value of the
left-hand side of Eq.~(\ref{noncomm}) is $\langle 0|[x^\mu, x^\nu] |0\rangle
= \langle 0|e^{-i\omega M}[x^\mu, x^\nu] e^{i\omega M} |0\rangle = \langle
0|[{x^{\prime}}^\mu, {x^{\prime}}^\nu] |0\rangle$, where ${x^{\prime}}%
=e^{-i\omega M}x\, e^{i\omega M}$ is the transformed to another Lorentz
frame operator-valued coordinate. The vacuum expectation value of the
right-hand side is $\langle 0|\theta^{\mu\nu}|0\rangle=\theta^{\mu\nu}$. For
these results of vacuum averaging to be compatible, $\theta^{\mu\nu}$ must
be Lorentz invariant, that is, this tensor must take the same numerical
values in every Lorentz frame. However, there is no Lorentz-invariant
antisymmetric second rank tensor in four-dimensional worlds.}. It has long
been known that a system is invariant under a given group of symmetry if and
only if the ground state of this system (vacuum) enjoys the property of this
symmetry\footnote{Sidney Coleman proved the impressively sounding theorem: 
``The invariance of
the vacuum is the invariance of world'' \cite{Coleman}.}.

The possibility for Lorentz violation is one of the currently hot problems
in different models of infrared-modified gravities \cite{Rubakov}.

Even a cursory examination of the mentioned surveys \cite{Mattingly},
\cite{Jacobson}, \cite{Douglas}, and \cite{Rubakov} gives a good idea of a large industry for
exploring a possible breakdown of Lorentz invariance, which arose recently.
The Lorentz violation problem is of concern in hundreds of papers which take
advantage of a full-fledged mathematical machinery stemming from many
branches of theoretical physics such as string theory, alternative theories
of gravitation, the early universe cosmology, brane worlds, loop quantum
gravity, etc.

On the other hand, it is rarely indeed that the conceptual aspect of this
problem is addressed. Needless to say that Lorentz invariance is a pillar of
the whole of modern physics. Therefore, if one poses the question as to
whether this invariance is broken, then it might be well to gain insight
into how much the geometry of Minkowski space is justified from the
operational and logical standpoints. An analysis of this kind might be of
utility, in particular to the question of whether the sought effects of
Lorentz violation are of order of $E/M_{\mathrm{P}}$, where $E$ is a
characteristic energy scale in observed processes, and $M_{\mathrm{P}}=\left(%
{\hbar c/{G}}\right)^{1/2}\simeq 1.2\cdot 10^{19}$ GeV the Planck mass, or
the actual suppression of these effects is much greater. Curiously, while on
the subject of these effects, our situation bears some resemblance to the
situation at the beginning of the 20th century when researchers tried to
understand whether the negative results of experiments for recording effects
of motion with respect to the luminiferous aether imply their suppression by
the factor of $v^2/c^2$ or there comes a point where the physical paradigm
must be changed completely.

Now the problem is to discriminate the basic facts, clearly formulated
suppositions and conventions, together with logical links between them, from
historical stratifications and personal psychological biases. This would
enable outlining the system of notions which are \textit{actually} at the
heart of special relativity. The paper is aimed at solving this problem.

\section{Inertial frames of reference}

\label{inertial} The essence of special relativity can be expressed, in the
spirit of Minkowski, in a single phrase: ``{Space and time are fused into
four-dimensional spacetime which is described by pseudoeuclidean geometry}%
.'' Mathematically, our physical spacetime is modeled on a four-dimensional 
\textit{flat} manifold, Minkowski space ${\mathbb R_{1,3}}$. Flat manifolds can be
naturally parametrized by {Cartesian} coordinates. In the language of
experimental physics, the use of Cartesian coordinates means that all
measurements of space and time are performed in {inertial} frames of
reference. So, before we proceed further, an effort must be made to refine
the notion of {inertial} frame of reference.

We first become aware of this notion in middle school where we gain an
impression of inertial frames by a series of simple comparative examples: a
carriage which is affected by pits and bumps---a car moving along a highway,
a carousel---a stationary school laboratory---a spaceship in its weightlessness flight.
In the studentship season we return to this notion at a higher level when we
ponder over the question: ``Why did Newton accentuate the statement that a
free body continues in its state of rest or uniform motion in a straight
line as a separate law even if the state of nonaccelerating motion is an
obvious consequence of Newton's second law in the absence of external
forces?'' At this juncture, our tutor reminds us that a free body moves at a
constant rate with respect to an \textit{inertial} frame of reference, while
the definition of this notion has not so far been given. In fact, the
formulation of Newton's first law is tantamount to defining the notion of
inertial frame.

This view is represented in many Western courses on General Physics and in
some texts on Theoretical Physics \cite{Barut}, \cite{Synge}. According to
this view, Newton's first law is to be thought of as the statement: ``There
is at least one frame of reference in which free particles move uniformly
along straight lines. Every frame which has a uniform motion of translation
relative to this frame is also an inertial frame.'' The Springer \textit{%
Encyclopedia of Physics} \cite{encyclop} gives essentially the same
definition of inertial frames. This treatment is held to be
recognized, so that, in the following, we will refer to this concept of the
inertial property as the \textit{orthodox} concept.

The surprising thing is that many of up-to-date textbooks which cover
mechanics as a branch of Theoretical Physics deign no elucidation of the
term ``inertial frame''. The authors of these books pretend that the reader
is already well aware of this notion (as, say, it happens with elementary
algebra), and hence it is unnecessary to expend time in recalling the
definitions. The higher the level of the presentation, the much more likely
that the author will refrain from specifics in this matter.

The reason for this ``restraint'' is that the orthodox concept of the
inertial property is not blameless in the \textit{logical} respect. Based on
this concept, we envision that a \textit{free} particle moves \textit{%
uniformly} along a \textit{straight line} in an inertial frame of reference,
assuming tacitly that we have at our disposal appropriate measurement
procedures and criteria for deciding whether a given particle is indeed
``free'', its motion is ``uniform'', and its path is ``rectilinear''.
Strictly speaking, none of these qualities is possible to verify {%
independently} of one another. The naive idea that ``a particle is almost
free if it is well off other particles'' is valid as a first approximation
until we recall \textit{self-interaction}. In fact, every particle is a
source of some field. Hence, the particle feels a back reaction of this
field, which is nonvanishing even if this particle is in a region of space
remote from other matter. The question of whether Newton's first law governs
the behavior of a self-interacting particle remains open. Experiment seems
to point clearly that the behavior of every sufficiently isolated object
obeys Newton's first law. Nevertheless, the problem is not quite trivial. It
may be that our observations are too rough or insufficiently purposeful to
be capable of grasping an anomalous behavior of self-interacting particles.
On the other hand, it seems plausible that recent measurements of redshifts
for distant supernovae \cite{Riess} make it clear that the universe as a
whole moves much differently than a free Galilean particle\footnote{%
In some cosmological scenarios, the universe is represented by a 3-brane (an
extended three-dimensional object with a particular dynamics) which is
embedded in a higher-dimensional space. Such a 3-brane generates the
gravitation field and is subjected to its back reaction. Therefore, the
3-brane does not necessarily follow the Galilean course even though the
other cosmological object, if they exist, are infinitely distant from the
3-brane. The experimentally recorded fact of accelerated recession of
galaxies may well be the manifestation of \textit{self-accelerated} motion
of the brane \cite{Carena}. This conjecture is an alternative to the
wide-spread belief that the accelerated expansion of the universe must be
ascribed to antigravitation related to the so-called dark energy \cite{Pad}.}%
. The physical notion of \textit{rectilinearity} can be modeled in
geometrical optics with the use of light rays. However, light propagates
along a straight line only in an inertial frame of reference. As to the
notion of \textit{uniform} motion, quite apparently, it cannot be defined
unless an inertial frame of reference is previously fixed because spacetime
measurements are sensitive to whether they are made in inertial or
noninertial frames. This leads to a vicious circle, that is, a notion $A$ is
defined by the notions $B$ and $C$ whose own sense is intangible unless they
are referred to notion $A$.

Is it possible to escape from this circle? Many people take this affair to
be inevitable and typical for any physical discipline which yet should not
discourage us because to define the fundamental notions and principles
individually and verify them separately is ill advised; only a {system}
analysis of these notions and principles and verification of them in their
unity is a distinct possibility. As Arnold Sommerfeld put it, the exposition
of electrodynamics must not begin with precise definitions because ``these
definitions are to be derived from a close inspection of their
interdependence in the \textit{basic equations of the theory} tested by
experiment'' \cite{sommerfeld}. A more developed motivation of this attitude
is given by Charles Misner, Kip Thorne, and John Wheeler \cite{Wheeler}: ``%
\textit{Here and elsewhere in science, as stressed not least by Henri
Poincar\'e, that view is out of date which used to say, <<Define your terms
before you proceed.>> All the laws and theories of physics, including the
Lorentz force law, have this deep and subtle character, that they both
define the concepts they use (here $\mathbf{B}$ and $\mathbf{E}$) and make
statements about these concepts. Contrariwise, the absence of some body of
theory, law, and principle deprives one of the means properly to define or
even to use concepts.}''

At times it may be that people are under the impression that avoiding the
troubles with the definition of inertial frames in the orthodox approach is not
unduly difficult if we invoke ``up-to-date'' theoretical tools. A case in
point is the Lev Landau and Eugeni{\u\i} Lifshitz' book \cite{L-L} which
holds: ``It is found, however, that a frame of reference can always be
chosen in which space is homogeneous and isotropic and time is homogeneous.
This is called an \textit{inertial frame}.'' In fact, the homogeneity and
isotropy are absolute (immanent) properties of space and time. Their
outwardness (if they exist) is unrelated to the use of inertial frames of
reference. In modern theoretical physics, spacetime is thought of as a
smooth manifold with postulated topological, metric, affine, and other
geometrical properties, which do not depend on a means of their description 
\cite{Hawking}. An appropriate choice of the frame of reference can only aid
in uncovering these properties, display them explicitly. On the other hand,
the use of noninertial frames in its own does not imply that the homogeneity
and isotropy of space and homogeneity of time are broken. The authors of the
famous \textit{Course of Theoretical Physics} were most likely well aware of
this fact\footnote{%
Curiously enough, the preceding paragraph (p.~5) reads: ``If we were to
choose an arbitrary frame of reference, space would be inhomogeneous and
anisotropic. This means that, even if a body interacted with no other
bodies, its various positions in space and its different orientations would
not be mechanically equivalent. The same would in general be true of time,
which would likewise be inhomogeneous; that is, different instants would not
be equivalent. Such properties of space and time would evidently complicate
the description of mechanical phenomena.''}. It is quite disappointing,
then, to see the negligence of the above formulation. Every so often the
reader swallows the text of the authoritative theorists uncritically%
\footnote{%
The first version of \textit{Mechanics}, which was written by Landau together
with L. Pyatigorsky, was published in 1940. In his 1946 review of this book,
Vladimir Fock noted: ``The considerations that precede the introduction of
the inertial frame are unclear, and it is hardly probable that the very
definition of such a frame as that <<firmly attached to freely moving
bodies>> (p. 17) is correct. A frame attached to a freely moving rotating
projectile comes within this definition'' \cite{Fock}. In the next 1957
edition of \textit{Mechanics}, which was written together with Lifshitz, the
definition of the inertial frame was altered. Just that altered definition is 
cited
above. In all subsequent editions the definition has remained unchanged,
and has survived, without any reader's protest, for more than 50 years!}.

Adopting the orthodox concept of the inertial property, we have to confess
frankly that the inertial property is taken as a physical quantity that
defies \textit{operational} definition. We must abandon any attempt to bring
the extent to which a given frame deviates from an inertial frame under
experimental control\footnote{%
For a discussion of this issue and further references see the essay by Jayant Narlikar 
\cite{Narlikar}.}. One may adduce a chain of the following propitiatory
arguments: If we need a ``standard'' inertial frame, then it can always be
chosen from theoretical considerations. To illustrate, one can use the
center of mass of the solar system with coordinate axes directed to distant
stars. In most of engineering problems, it is sufficient to use a frame of
reference rigidly attached to the Earth. If the problem requires a
high-accuracy treatment, then we use a frame related to the center of the
Earth and coordinate axes directed to distant stars. Small deviations from
the inertial property in this frame can be estimated by the indirect route
through introducing theoretical corrections for centrifugal and
gravitational effects. In actual practice, the measurement of these
deviations with physical measuring tools was yet unnecessary. Is it
possibile in principle to measure the inertial property variation in a
direct way? This statement of the problem was, to our knowledge, never
considered.

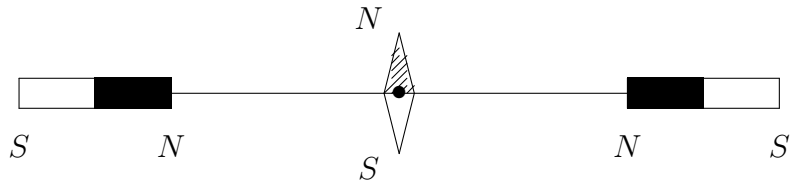
\begin{figure}[tbh]
\begin{center}
\unitlength=1mm \special{em:linewidth 0.4pt} \linethickness{0.4pt} 
\begin{picture}(110.00,23.00)
\emline{10.00}{17.00}{1}{10.00}{13.00}{2}
\emline{10.00}{13.00}{3}{20.00}{13.00}{4}
\emline{20.00}{13.00}{5}{20.00}{17.00}{6}
\emline{20.00}{17.00}{7}{10.00}{17.00}{8}
\emline{20.00}{17.00}{9}{30.00}{17.00}{10}
\emline{30.00}{17.00}{11}{30.00}{13.00}{12}
\emline{30.00}{13.00}{13}{20.00}{13.00}{14}
\put(30.00,13.00){\rule{-10.00\unitlength}{4.00\unitlength}}
\emline{90.00}{17.00}{15}{90.00}{13.00}{16}
\emline{90.00}{13.00}{17}{100.00}{13.00}{18}
\emline{100.00}{13.00}{19}{100.00}{17.00}{20}
\emline{100.00}{17.00}{21}{90.00}{17.00}{22}
\emline{100.00}{17.00}{23}{110.00}{17.00}{24}
\emline{110.00}{17.00}{25}{110.00}{13.00}{26}
\emline{110.00}{13.00}{27}{100.00}{13.00}{28}
\put(90.00,13.00){\rule{10.00\unitlength}{4.00\unitlength}}
\put(20.00,17.00){\rule{10.00\unitlength}{0.00\unitlength}}
\put(20.00,13.00){\rule{10.00\unitlength}{4.00\unitlength}}
\emline{30.00}{15.00}{29}{90.00}{15.00}{30}
\put(60.00,15.00){\makebox(0,0)[cc]{$\bullet$}}
\put(30.00,8.00){\makebox(0,0)[cc]{$N$}}
\put(10.00,8.00){\makebox(0,0)[cc]{$S$}}
\put(90.00,8.00){\makebox(0,0)[cc]{$N$}}
\put(110.00,8.00){\makebox(0,0)[cc]{$S$}}
\put(56.00,25.00){\makebox(0,0)[cc]{$N$}}
\put(56.00,5.00){\makebox(0,0)[cc]{$S$}}
\emline{58.00}{15.00}{31}{60.00}{23.00}{32}
\emline{60.00}{23.00}{33}{62.00}{15.00}{34}
\emline{62.00}{15.00}{35}{60.00}{7.00}{36}
\emline{60.00}{7.00}{37}{58.00}{15.00}{38}
\emline{58.00}{15.00}{39}{61.00}{18.00}{40}
\emline{59.00}{17.00}{41}{61.00}{19.00}{42}
\emline{59.00}{18.00}{43}{61.00}{20.00}{44}
\emline{59.00}{19.00}{45}{60.00}{20.00}{46}
\emline{59.00}{15.00}{47}{61.00}{17.00}{48}
\emline{61.00}{15.00}{49}{62.00}{16.00}{50}
\emline{60.00}{15.00}{49}{61.00}{16.00}{50}
\emline{60.00}{21.00}{32}{59.00}{20.00}{34}
\end{picture}
\end{center}
\caption{Magnetic needle in the state of unstable equilibrium}
\label{paths}
\end{figure}

Another concept of the inertial property is proposed in Ref.~\cite{k06}. A
key idea here is to make an argument involving the notion of \textit{%
unstable equilibrium}. Intuition suggests that states of unstable
equilibrium are maintained only in inertial frames, because shocks and blows
associated with accelerated motions of noninertial frames prevent unstable
systems from being balanced. We thus come to a simple operational criterion
for distinguishing between inertial and noninertial frames based on the
capability of inertial frames for preserving unstable equilibrations. As an
example, a suitable device which controls the inertial property is shown in
Figure~\ref{paths}. 
A magnetic needle is installed halfway between north poles of two identical
static magnets on the axis along which the magnets are lined up. A state of
unstable equilibrium is attained when the needle is perpendicular to this
axis. The magnets are mounted rigidly to the frame and their separation is
fixed. A small perturbation will suffice for the needle to swing through $%
+90^{\circ }$ or $-90^{\circ }$, so that its resulting direction is aligned
with the magnet axis or is opposed to it. Such a swing signals that some
violation of the inertial property occurs, no matter how small this
violation may be. (If small shocks of arbitrary directions are to be
detected, three such devices with noncomplanar axes are in fact required.)  
The advantage of this concept is that it is appropriate for distinguishing
not only between idealized inertial and noninertial frames in {thought
experiments} but also between real frames which are inertial frames only
approximately. Furthermore, this opens a simple and natural way for
rendering the inertness a \textit{measured} quantity. To do this requires a
testing system which is stable against small perturbations but unstable
against perturbations above some finite threshold. If we employ testing
systems of a definite type, then the inertness measure can be expressed in
terms of the threshold magnitude. For example, taking the testing system
shown in Figure~\ref{paths}, use could be made of the frictional bond
(static friction) of the magnetic needle axle. Of course, for a practical 
implementation of this measuring procedure, we need a device which is
calibrated with using testing systems of different types.

\begin{figure}[htb]
\begin{center}
\unitlength=1.00mm \special{em:linewidth 0.4pt} \linethickness{0.4pt} 
\begin{picture}(80.00,30.00)
\bezier{180}(10.00,10.00)(20.00,30.00)(30.00,10.00)
\put(20.00,21.00){\makebox(0,0)[cc]{$\circ$}}
\emline{40.00}{15.00}{1}{80.00}{15.00}{2}
\put(60.00,16.00){\makebox(0,0)[cc]{$\circ$}}
\end{picture}
\end{center}
\caption{Particle in the states of unstable and neutral equilibrium}
\label{bounbary}
\end{figure}
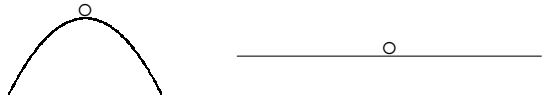

It is particularly remarkable that the orthodox concept of the inertial
property can also be understood in the context of equilibrium states. To see
this, we note that free particles in an inertial frame are objects in the
state of \textit{neutral equilibrium}. Two approaches to the definition of
inertial frames are illustrated schematically in Figure~\ref{bounbary} which
depicts a particle at the top of a potential hill and that in a constant
potential.

Let us compare these approaches. The definition based on the notion of
unstable equilibrium is introduced prior to geometry, that is, using only a
kind of yes-no decision: if the state of unstable equilibrium remains
unchanged, then the frame is inertial, otherwise the frame is noninertial.
On the other hand, the geometry is important for the orthodox approach which
relies on the notion of neutral equilibrium. Indeed, let a particle be in
the state of neutral equilibrium. This is another way of stating that the
particle continues in its state of rest or \textit{uniform} motion. However,
the motion is seen to be uniform as the time read from the clock of a
definite construction whereas it is no longer uniform if we use another time
scale (for more detail, see Section 4). Therefore, there is no way to
discriminate between the states of perfect and flawed neutral equilibrium
unless the time-keeping is fixed. More generally, the notion of neutral
equilibrium is meaningful only in the context of a well-defined geometry
with specification of the experimental procedure for probing this geometry.

The result of yes-no decision is clearly visible. Mathematically, the yes-no
alternative translates into broken and unbroken discrete symmetries of
testing devices. For example, this translation, as applied to the device
shown in Figure~\ref{paths}, is either two perpendicular symmetry axes
referring to the state of unstable equilibrium or a single symmetry axis
referring to the case that the equilibrium is disturbed\footnote{%
With this observation, it is possible to define the concept of inertial
frames in a mathematically rigorous, axiomatic manner. We give a brief sketch of this
definition by saying that, for some systems, there are two kinds of states, {%
stable} and {unstable}, which differ drastically in their symmetry
properties. Now, the criterion for discriminating inertial and noninertial
frames reads: unstable states (with their unique discrete symmetry
properties) exist only in inertial frames.}. As to a similar symbolization
of the state of neutral equilibrium, this state exhibits invariance only
under {continuous} groups such as the translation group (it is this symmetry
group which accounts for the fact that space is homogeneous). If it is granted that a particle is placed in a
box, then the translation invariance is broken. However, this symmetry
breakdown bears no relation to rendering the inertial frame noninertial
because the very presence of the walls has no effect on the fact that a
particle in the box is in the state of neutral equilibrium. In a sense this
proposition is that inverse of the statement given above. To sum up, letting
the frame to be accelerated, this does not imply that space is
inhomogeneous, and conversely, letting space homogeneity to be broken, this
does not lead automatically to changing the frame from inertial into
noninertial.

In the subsequent discussion, we will always keep in mind the definition of
inertial frames of reference given in \cite{k06}.

\section{Space measurements}

\label{space} The fundamental premise of relativistic physics, the existence
of \textit{maximal} velocity of movers, is a simple idea, and yet a
satisfactory, clear-cut explication of this idea is difficult to extract
from the current literature.

Looking at waves of different nature (optical, sonic, etc.), which expand
from some point of emission $O$, one can compare their propagation rate even
without knowledge of the numerical values of their velocities. This is in
the same spirit as an ancient allegory: Achilles is prompter than a
tortoise, a dart is prompter than Achilles, a thunderbolt is prompter than
darts, etc. Mathematically, we have a chain $t<A<d<\ldots$ The ordering is
verified by inspection, and has no need of particular velocity scale. One
may suppose that there is an {upper bound} for propagation rates. (In fact,
this supposition is well verified experimentally.) A light wave runs down
any mover which left $O$ before the light emission, except for another light
wave emitted in $O$. The existence of the highest propagation
rate---universally referred to as the \textit{speed of light}, is a central
tenet of relativity. It is generally believed that the fundamental
interactions of nature propagate at the speed of light. This should be
compared with pre-relativistic physics in which arbitrarily high velocities
of bodies are allowable, and interactions (such as Newtonian gravitation)
are instantaneous.

We thus proceed from the assumption that a linearly ordered set 
\begin{equation}
t<A<d<\cdots  \label{sequence}
\end{equation}
is bounded above. By the presence of a supremum is meant the condition ``%
\textit{light does not overtake light, nothing can overtake light}''.
Clearly this condition is introduced prior to geometry. However, all is in
readiness to establishing the geometry. Given a clock and a radar ranging
device, we have actually everything required for probing the geometry of the
real world. Indeed, we can measure the distance between two points by the
time that light takes to traverse it. Taking the speed of light to be unity,
we find the distance we seek to be half the round-trip span between these
points.

Why is radar location preferable to using yardsticks? In practice, it is not
always possible to lay out a grid of the yardsticks. Take, for example,
cosmic measurements. But what is more important is that, when moving, the
yardstick is purported to preserve size and shape, at least in the absence
of stresses, temperature variations, and influences of electromagnetic and
gravitational fields. However, the self-congruence of a transported
yardstick has yet to be proved. From the pedestrian standpoint, it seems
unnecessary to ponder here, but, at the fundamental level, the
self-congruence problem is very serious\footnote{%
For an extended discussion of this problem see Gr\"unbaum \cite{Grunbaum}.}.
By contrast, with the radar-location approach, the size and shape of a given
rod can be {verified} by continuously sounding its ends. Furthermore, the
notion of {rectilinearity}, stemming from geometrical optics in which light
rays travel along straight lines is an integral part of the radar-location
approach.

Now, armed with a well defined and controllable standard for measuring
lengths and for determining what a straight line is, we are allowed the
option of using either yardstick or radar for a particular measurement.

Subsequent to choosing the length scale, we can define the velocity scale
[in other words, we can endow the set (\ref{sequence}) with a metrical
structure]. In the relativistic context, it is natural to take the
convention that the speed of light {in vacuum} is 1, and measure space and
time intervals in the same units, say, in seconds. Then the velocities of an
ordinary mover are dimensionless numbers whose magnitudes are no more than 1%
\footnote{%
Conceivably, the mathematically inclined reader will notice that the
mechanism for grafting a metric structure onto the linearly ordered and
limited from above set (\ref{sequence}) is similar to the mechanism for
endowing the real axis 
$\mathbb R$
with the natural metric in response to distinguishing a subset of ``upper
bounds'', the set of integers 
$\mathbb Z$, among this linearly ordered set.}.

\section{Time measurements}

\label{time}

The final component of the spacetime measuring succession based on the
radar-location method is time measurements. In giving the rationale of this
method, the major part of the conceptual burden falls on just this component.

In order to compare time durations of various physical phenomena, including
spatially separated ones, it is necessary to choose a time measure. One
takes a {``regularly recurrent'' process} as a standard, and the period of
this process $T$ as a unit of time. Thereafter the time-keeping is reduced
to simply counting the number $n$ of periods $T$ contained in the measured
interval.

The question on the measurement accuracy in this clocking can be raised only
in the case that $n\gg 1$. However, our concern here is on the matter of
principle rather than the limitation on the area of application. This
clocking is based on the indistinct notion of \textit{regularly recurrent}
processes, and hence the measurement result turns out to be somewhat
arbitrary.

A modern experimentalist may be furnished with a large amount of different
types of cyclic mechanisms operating on \textit{various} time, and hence
counting different number of seconds for the same time span. For example, he
may provide himself with a set of clocks of every possible construction,
beginning with an old cuckoo-clock and ending with recent models of Rolex
watches with quartz movements. For reference, in ancient time, this set of
chronometers would involve only sun-dials, water-clocks, and sand-glasses.
Devices that transform accelerated motions of a falling body into uniform
rotations of a minute hand did not appear until Middle Ages. In 1583,
Galilei discovered the fact that small swings of pendulum are isochronous
(that is, the period of small oscillations of a fixed-length pendulum is
amplitude-independent) with the help of beating of his pulse. In 1657,
Huygens took out the patent for pendulum clock\footnote{%
The story of the pendulum clock invention can be found in the
Gindikin book \cite{Gindikin}.}.

The problem would be solved immediately if a certain time scale were
distinguished among other scales.

Newton believed that there exists \textit{Absolute, True, Mathematical Time} 
$t$, which in itself, and from its own nature flows equably without regard
to any thing external, and by another name is called Duration. He contrasted
Absolute Time $t$ with \textit{Relative, Apparent, Common Time} which, in
his opinion, is a sensible and external (whether accurate or unequable)
measure of Duration by the means of motion, which is commonly used instead
of True time; such as an Hour, a Day, a Month, a Year...

However, Newton did not define a procedure for checking the rate of a given
cyclic mechanism against the flow of absolute time $t$. Therefore, the
Newtonian absolute time is a \textit{metaphysical} quantity which eludes 
\textit{operational} definition. It is well known that Newton did away with
any {metaphysics} in his physics research. The rare exceptions to this rule
are likely to be evidence that these notions are very deep. From a pragmatic
point of view, it is reasonable to define such an evolution parameter $t$
which would ensure the \textit{simplest} formulation of the laws of physics,
in particular, the motion of a free particle must be uniform with respect to
the frame in which time is measured in units of the defined scale. If we
use, instead, another parameter of evolution $\tau$ which increase
monotonically with $t$, say, $\tau=t_0\ln\left(1+t/t_0\right)$, then the
particle's velocity is $\tau$-dependent, and so the behavior of free
particles is formally more complex.

Carl Neumann was the first to realize in 1870 that the laws of mechanics
assume their simplest form if the Newtonian absolute time is used, otherwise
these laws become more complicated \cite{Neumann}.

Now the reader may form the impression that the desired procedure for
checking the rate of a given clock against the flow of absolute time is
ultimately clarified. The famous Misner--Thorn--Wheeler's burden \cite%
{Wheeler} comes to mind: ``Time is defined so that motion looks simple''.
Unfortunately, this impression is wrong. Indeed, if time is defined so that
the motion of a free particle looks uniform, then Newton's first law is a
mere tautology. However, we would like to have a system of conventions which
provides a way of verifying fundamental physical laws independently of one
another.

One is then inclined to consider a weaker requirement for an appropriate
time scale in a given inertial frame of reference. The requirement reads:
with such a scale, uniform motion of every \textit{inertial frame} with
respect to this frame must be attained. However, a reservation is to be made
that the set of frames under examination need not include all free particles%
\footnote{%
It may be argued that, when on the subject of mechanical objects, the
discrimination between ``frames of reference'' and ``free particles'' seems
rather strained. In the subsequent discussion, we restrict our attention to
macroscopic nonrotating platforms, equipped with measuring devices for space
and time measurements if we are to consider an embodiment of frames of
reference. In the next section, we will discuss at some length why such
objects may be regarded as distinguished among all conceivable and
theoretically allowable variety of mechanical objects.}. It is natural to
call this scale \textit{standard} or \textit{laboratory} scale \cite{k06}.
The availability of the criterion for distinguishing between inertial and
noninertial frames based on yes-no decision, without reference to the
spacetime geometry, makes this definition logically consistent.

Now let a stationary observer be equipped with a set of different types of
clocks operating on various time and a radar ranging device. Suppose that
another inertial frame has a uniform motion of translation relative to him
along the $x$-axis. If the observer's time is read from a clock with such
scale $t$, then the motion is uniform: 
\begin{equation}
\frac{dx}{dt}=V=\mathrm{const}.  \label{dz-over-dt=v=const}
\end{equation}
By comparison, if he takes another scale $\tau$, then the inertial frame
executes a nonuniform motion, that is, its velocity varies with $\tau$: 
\begin{equation}
\frac{dx}{d\tau}={v}(\tau).  \label{dz-over-dtau}
\end{equation}
Let the function $v={v}(\tau)$ be known. Then the scale $t$ readily regains, 
\begin{equation}
t=f(\tau).  \label{t=F(tau)}
\end{equation}
Indeed, combining (\ref{dz-over-dtau}) and (\ref{t=F(tau)}) gives 
\begin{equation}
\frac{dx}{dt}=\frac{dx}{d\tau}\,\frac{d\tau}{dt}={v}\,\frac{{1}}{f^{\prime}}%
\,.  \label{dz-over-dt=dz-over-dtau-1-over-F}
\end{equation}
This, together with (\ref{dz-over-dt=v=const}), shows that 
\begin{equation}
f(\tau)=\frac{1}{V}\int^\tau d\xi\, {v}(\xi)+T_0,
\label{F=1-over-v-int-dz-over-dtau}
\end{equation}
where $T_0$ is an arbitrary integration constant. Expressions (\ref%
{dz-over-dtau}) and (\ref{F=1-over-v-int-dz-over-dtau}) suggest the way for
rearranging the given chronometer to yield the standard scale $t$.

The standard scale $t$ is defined up to a linear transformation 
\begin{equation}
t^{\prime}=kt+ T_0.  \label{rescaling}
\end{equation}
In other words, the definition of $t$ leaves room for a global change of the
scale $k$ (say, hours may be termed minutes---which is of course a matter of
convention), and shifting the zero of time by $T_0$. The condition of
uniform motion (\ref{dz-over-dt=v=const}) is unaffected by the linear
transformation (\ref{rescaling}) because its action is equivalent to
replacing the constant ${V}$ with another constant ${V/k}$ in Eq.~(\ref%
{dz-over-dt=v=const}) The uniformity of motion of the tested inertial frame
is violated if and only if $f(\tau)$ in Eq.~(\ref%
{F=1-over-v-int-dz-over-dtau}) is a \textit{nonlinear} function.

It may be worth pointing out that clocks which read the standard time rate
should not be regarded as \textit{true}, or \textit{accurate}. However, this
time rate may be judged \textit{most convenient}, and the associated history
with the parameter of evolution $t$ \textit{much simpler} than other
histories.

At present, the closest fit to the standard time rate $t$ shows atomic
clocks. According to the fundamentals of quantum mechanics (which will be
not touched upon here) the Compton wave length of an elementary particle $%
\lambda$, and the corresponding wave period $T=\lambda/c$ remain constant if
the measuring system is calibrated with the aid of the standard time scale.

It might seem that a ``perfect periodicity'' can be attained if light rays
are sandwiched between two parallel mirrors. This device is often mentioned
in the literature, see, e.~g., \cite{Wheeler}. To measure time one simply
counts successive reflections of this light shuttle. A virtue of this light
clock is that spatial and temporal measurements are assembled into a single
process of sending light flashes and receiving their echoes. But a closer
look at this timekeeping reveals a serious flaw: the mirrors must be
separated by a \textit{fixed} distance, which is a problem. In the absence
of reliable space control procedure different from radar sounding, this
chronometry would make the definition circular. This problem is in effect
the same as the problem on preserving the self-congruence of a yardstick.

We will therefore content ourselves with the very existence of a clock, such
as atomic clocks, whose readings conform to the requirement that relative
motions of inertial frames be uniform. In the subsequent discussion we
assume that inertial frames are equipped with clocks of this type.

\section{The basic tenets of special relativity}

\label{esse} Let us recall, in a few words, nodal points in the history of
formation of the geometric principles of special relativity. Analyzing the
concepts of space and time, and guided by the experimental situation at the
turn of the century, Poincar\'e \cite{Poinc1898}---\cite{Poinc1904}
concluded that no motion with respect to the aether (or, what is the same,
with respect to Newton's Absolute Space) is detectable. He formulated the 
\textit{principle of relativity}, which, coupled with the idea of a \textit{%
maximum signal velocity}, offered a clear view that the \textit{simultaneity}
of spatially separated events is a matter of \textit{convention}. Einstein 
\cite{Einstein1} discarded the concept of an aether, and proceeded from two
postulates: (1) the laws of physics are the same in all inertial frames, and
(2) the speed of light in vacuum is constant. He proposed a convention for
simultaneity of separated points, now known as the standard synchrony, which
provided a means of simple and natural deriving the Lorentz transformation.
This transformation was earlier derived by Larmor \cite{Larmor} and Lorentz 
\cite{Lorentz} on the hypothesis that any effect of the motion through the
aether is unobservable. Poincar\'e \cite{Poinc1905} showed that the set of
all Lorentz transformations, combined with space rotations, constitutes a
group, to which he gave the name the Lorentz group. A pseudoeuclidean
spacetime metric was first introduced by Poincar\'e \cite{Poinc1906}.
Minkowski \cite{Minkowski} unified space and time into an indivisible entity
which he called ``the world'' and described it in terms of {four-dimensional
pseudoeulidean geometry}\footnote{%
For more historical details and further references see Whittaker \cite%
{Whittaker1953}, and Pais \cite{Pais}.}.

With the novel understanding of inertial frames, we should be aware of how
this may reflect on the formulation and content of the base of special
relativity. Let us begin with the {principle of relativity}.

The definition of inertial frames given in \cite{k06} does not claim from
the very beginning that there exists a kinematical link between all such
frames. The inertial property of each frame is examined individually---from
a ``yes-no'' decision which is implemented in the given subject frame. The
kinematical link is produced further as a matter of \textit{convention}: we
agree to regard a definite time scale as the standard scale.

By contrast, if we proceed from the orthodox concept of the inertial
property, then the matter of convention is larger, the dynamical law for
free particles. Such a convention, however, is too restrictive. There is
little point in the \textit{{a} priori} requirement that every free object,
rather than the inertial frames alone, move uniformly along straight lines
on condition that the standard time scale has been used.

What are objects implied when we make the reservation that, in the
absence of external forces, they do not necessarily move along
straight world lines, that is,  their
behavior is non-Galilean? Whether is this behavior
consistent with the law of energy conservation? 
Why then should what is meant by a frame of reference behave in
the Galilean fashion?

Non-Galilean regimes of motion are in principle possible. Their existence is
consistent with every fundamental physical law in the scope of special 
relativity. In
Ref.~\cite{k06}, four types of objects that are capable of evolving in
non-Galilean regimes are described. The simplest one is a
particle living in a realm with \textit{discrete time}. It is supposed that
this particle is governed by Newton's second law which,
however, takes the form not of the usual {differential} equation $m{\dot{%
\mathbf{v}}}=\mathbf{f}$ but of the difference equation 
\begin{equation}
m\,\frac{\mathbf{v}(t+{\ell})-\mathbf{v}(t)}{\ell}=\mathbf{f},
\label{difference-Newton}
\end{equation}
where $\ell$ is a ``quantum of time''. The general solution to Eq.~(\ref%
{difference-Newton}) with $\mathbf{f}=\mathbf{0%
}$ is 
\begin{equation}
\mathbf{v}(t)=\sum_{n=0}^\infty\,\mathbf{C}_n\cos\left(\frac{2\pi n}{\ell}t
+\varphi_n\right), \quad  \label{difference-Newton-solution}
\end{equation}
where $\mathbf{C}_n$ and $\varphi_n$ are integration constants. This means
that the free particle executes a periodic motion, the so-called
zitterbewegung, back and forth along a straight line parallel to $\mathbf{C}_n$ for $%
n\ge 1$, and moves uniformly for $n=0$.

It seems likely that the only interval in nature which may be promoted to
the status of quantum of time is the Plank interval $t_{\mathrm{P}%
}=\left(\hbar G/c^5\right)^{\frac12}\approx 0,5\cdot 10^{-43}$ sec. Quite
apparently, the ``quantum'' $t_{\mathrm{P}}$ is tiny in comparison with spans
of real measurements. Therefore, if a macroscopic
platform to be used as a frame of reference executes a zitterbewegung with
the period $\ell$ of order of $t_{\mathrm{P}}$, this have
no effect on practical measurements in this frame.

In addition, there are two objects which are capable of executing 
zitterbewegungs. First, a particle with intrinsic angular momentum, \textit{%
spin}\footnote{%
For example, a vanishingly small gyroscope which appears in the Frenkel
model \cite{Frenkel}.}, and, second, a particle with the so-called \textit{%
rigid dynamics}, that is, a particle whose Lagrangian involves not only
coordinates and velocities but also higher derivatives\footnote{%
Such particles may occur quite naturally if one attempts to construct a
consistent Lagrangian for Maxwell's electrodynamics in spacetimes of
dimension greater than $D=4$. For example, for $D=6$, the Lagrangian should
depend on both velocities and accelerations of charged particles \cite%
{Kosyakov}--\cite{Yaremko}.}.

If we require that mechanical objects used as frames of reference have zero
angular momentum, and assume that the physical effects of noncompactified
higher dimensions are too small to affect the behavior of macroscopic
bodies, then it is reasonably safe to suggest that zitterbewegungs are
excluded from the number of non-Galilean regimes of motion allowable for
frames of reference in the absence of external forces.

The fourth type of non-Galilean objects comprises \textit{self-interacting
particles}. Two of them, covered in \cite{k06} at some length, are
self-interacting particles in the Maxwell--Lorentz electrodynamics and
classical Yang--Mills theory. In the absence of external forces, they 
either execute uniform motion in straight lines or
continually accelerate/decelerate with exponentially increasing/decreasing 
acceleration. A given
charged particle can move in either Galilean or self-accelerated regimes
according to which the asymptotic condition for this particle is realized in
the far future. An important point is that this regime remains the same
throughout the whole history of this particle. In other words, there exist
two classes of particles, so that the transition of a particle from one
class to another is forbidden. Our daily experience in relation to the
behavior of macroscopic bodies suggests that all belong to the class of
Galilean objects. It would not be appropriate here to analyze why the
experiment still failed to record non-Galilean objects with exponentially
increasing acceleration, even though their existence would not run counter
to the basic principles of physics\footnote{%
Up to now, there has been no serious discussion of this problem in the
literature. Some theorists do not see the problem altogether. Their
characteristic attitude towards a solution describing self-accelerated
motion is shown in the term ``pathological solution''. Many of them believe
that non-Galilean regimes of evolution are inconsistent with the law of
conservation of energy. It is shown in \cite{k06} that this belief is
erroneous. The energy balance is exactly obeyed. A key observation here is
that the energy of a \textit{dressed} particle need not be a positively
defined quantity.}. Within the limits of the present discussion, it will be
sufficient to take as a fact the existence of the large class of Galilean
objects suitable for realization in practice, of the idea of inertial frame
of reference.

As to the hypothesis of maximal velocity of signal propagation, an important
comment of it is also in order. The original sense of this hypothesis was
that the set (\ref{sequence}) has a supremum, or, in the physical language,
``{light does not overtake light, nothing can overtake light}''. We next
endowed the set (\ref{sequence}) with a metrical structure in such a way as
to choose the metric in its simplest form. For this to happen, the light
propagation should appear \textit{uniform} provided that the clock reads the 
\textit{standard} time rate. This may be interpreted as refining the
Einstein's postulate of the constancy of the velocity of light. Attention is
usually drawn to the condition that the velocity of light is the same in all
inertial frames\footnote{%
This condition is normally taken to be odd because whatever the motion of
the source of light, this leaves the velocity of light propagation
unaffected, whereas the naive common sense guides us to suppose that the
source velocity must be added to or subtracted from the wave front velocity
depending on whether the wave front is aligned with the source's direction
of motion or is opposed to it. However, on second thought, we conclude that
``common sense'' let us down. Until the principle of maximal velocity was
formulated, we may not assume that a rule for the addition of velocities is
already defined for elements of the set (\ref{sequence}). Once this
principle has been adopted, we are entitled to reason about affine structure
of spacetime, Cartesian coordinate systems, and the rule for the addition of
velocities. It transpires that, considering the collinear motion of a light
wave and its source and choosing the simplest, from the relativistic
standpoint, metric on the set (\ref{sequence}), the algebraic addition
applies not to the velocities $V$ but to the rapidities $\frac12\ln\left(%
\frac{1+V}{1-V}\right)$.}. But due regard must be given to another aspect of
this postulate, the \textit{uniformity} of light propagation.

\section{Conclusion}

\label{conclusion} Let us summarize our discussion. We have sought to
clarify whether it is possible to ``reconstruct'' in a coherent and
logically consistent way the basic geometrical properties of spacetime in
special relativity. (We use the term ``reconstruction'' in a conventional
sense. As was pointed out by Poincar\'e, the geometry by itself eludes
verification. What is to be verified is the totality of geometry and
physics. A change of geometric axioms can be accompanied by a suitable
modification of physical laws in such a way that the predictions of observed
phenomena are unchanged. Nevertheless, the entire theoretical scheme is not
indifferent to the choice of a particular geometry\footnote{%
As an illustration, we refer to the problem of the field generated by a
magnetic charge. This problem can be stated in two alternative geometric
settings. The vector potential due to a magnetic monopole $\mathbf{A}$ as
viewed in the usual Euclidean space is singular on a line that issues out of
the magnetic charge, the so-called Dirac string. On the other hand, the
vector potential $\mathbf{A}$ as viewed in a manifold which is obtained by
gluing together two Euclidean spaces is regular everywhere except for the
point where the magnetic charge is located.}.) Of course, our interest is
with the simplest physically justifiable version of spacetime.

Special relativity assumes that spacetime is described by the geometry of {%
Minkowski space} ${\mathbb R_{1,3}}$. It is generally believed that 
${\mathbb R_{1,3}}$ is the
best geometric framework for the great bulk of phenomena, excluding
situations in which strong gravitational fields are present. The reason for
this belief is threefold:

(i) the existence of \textit{inertial} frames,

(ii) the existence of a \textit{standard} time scale,

(iii) the existence of the \textit{highest} propagation rate, associated
with the {speed of light}; the \textit{uniformity} of light propagation.

An inertial observer, having clocks with the standard time scale, will
recognize {time as \textit{homogeneous}} in the sense that all instants are
equivalent. With the idea of the existence of {maximal velocity} for signal
propagation in mind, this observer equipped with a radar ranging device will
find {space \textit{homogeneous} and \textit{isotropic}} in the sense that
there are no privileged position in space and distinguished direction of
motion. These facts can serve as the natural basis for reconstructing the 
\textit{affine} structure of spacetime\footnote{In non-inertial 
frames, it is also possible to reconstruct the {affine} structure 
by the indirect route. Suppose we have experimental evidence of the laws of 
conservation
of momentum,  energy, and angular momentum. Then we take advantage of the 
inversion of Noether's first theorem \cite{Palmieri} to
discover invariance under the 
group of space and time translations and the group of space rotation, which is 
another way of stating that time is homogeneous and space is
homogeneous and isotropic.}.

To this must be added the principle of relativity for inertial frames of
reference to reconstruct the \textit{metrical}, pseudoeuclidean structure
of spacetime.

\section*{Acknowledgment}

This paper was written, in part, during a visit of one of the authors (BPK)
in the Autonomous University of Zacatecas. BPK thanks the University
Administration for the warm hospitality.


\begin{thebibliography}{99}
\bibitem{Mattingly} Mattingly, D. \textit{Living Rev. Relativity} \textbf{8}
5 (2005).

\bibitem{Jacobson} Jacobson, T., S. Liberati, and D. Mattingly \textit{Ann.
Phys.} \textbf{321} 150 (2006).

\bibitem{Seiberg} {Seiberg, N.,} and E. Witten \textit{J. High Energy Phys.}
\textbf{09} 032 (1999).

\bibitem{Douglas} {Douglas, M. R., and N. A. Nekrasov} \textit{Rev. Mod.
Phys.} \textbf{73} 977 (2001).

\bibitem{Wightman} Bogoliubov, N. N., A. A. Logunov, A. I. Oksak, and I. T.
Todorov \textit{General Principles of Quantum Field Theory} (Dordrecht:
Kluwer, 1990).

\bibitem{Coleman} {Coleman, S.} \textit{J. Math. Phys.} \textbf{7} 787
(1966).

\bibitem{Rubakov} Rubakov, V. A., and P. G. Tinyakov \textit{Phys. Usp.} 
\textbf{51} 759 (2008).

\bibitem{Barut} Barut, A. O. \textit{Electrodynamics and Classical Theory of
Fields and Particles} (New York: Colier--Macmillan, 1964). 2nd edition: New
York: Dover, 1980.

\bibitem{Synge} Synge, J. L. \textit{Relativity: The Special Theory}
(Amsterdam: North Holland, 1956).

\bibitem{encyclop} Bergman, P. G. \textit{The Special Theory of Relativity}
In: \textit{Encyclopedia of Physics} v. 4. Ed. by S. Fl{\"u}gge (Berlin:
Springer, 1962), pp. 109-202.

\bibitem{Riess} Riess, A. G. \textit{et al}. \textit{Astron. J}. \textbf{116}
1009 (1998); \textit{Astrophys.\ J}. \textbf{536} 62 (2000), \textbf{560} 49
(2001); Perlmutter, S. \textit{et al}. \textit{Astrophys.\ J}. \textbf{517}
565 (1999).

\bibitem{Carena} Carena, M., J. Lykken, M. Park, and J. Santiago \textit{%
Phys. Rev.}  \textbf{D 75} 026009 (2007).

\bibitem{Pad} Padmanabhan, T. \textit{Phys. Rep.}  \textbf{380} 235 (2003).

\bibitem{sommerfeld} Sommerfeld, A. \textit{Elektrodynamik} (Leipzig:
Akademische Verlasgsgesellschaft Geest {\&} Portig, 1949).

\bibitem{Wheeler} Misner, C. W., K. S. Thorne, and J. A. Wheeler \textit{%
Gravitation} (San Francisco: Freeman, 1973).

\bibitem{L-L} Landau, L. D., and E. M. Lifshitz \textit{Mechanics} Volume 1
of \textit{Course of Theoretical Physics} (Oxford: Pergamon, 1960). 2nd
edition 1969, 3rd edition 1976. Reprinted 1978, 1982, 1984, 1986, 1987,
1988, 1989, 1991, 1996, 1997, 1998, 1999, 2000. Translated from the 3rd
Russian edition of \textit{Mekhanika}, Moscow, Nauka, 1993.

\bibitem{Hawking} Hawking, S. W., and G. F. R. Ellis \textit{The Large Scale
of Spacetime} (Cambridge: Cambridge University Press, 1973).

\bibitem{Fock} {Fock, V. A.} \textit{Uspekhi Fizicheskikh Nauk} \textbf{28}
377 (1946) (in Russian).

\bibitem{Narlikar} Narlikar, J. V. \textit{Inertia and cosmology in
Einstein's relativity}. In: \textit{Astrofisica e Cosmologia Gravitazione
Quante e Relaivita} Negli sviluppi del pensiero scientifico di Albert
Einstein (Firenze: Giunti Barbera, 1979)

\bibitem{k06} Kosyakov, B. \textit{Introduction to the Classical Theory of
Particles and Fields} (Berlin: Springer, 2007).

\bibitem{Gindikin} Gindikin, S. \textit{Tales of Physicists and
Mathematicians} (Princeton: Princeton University Press, 2001).

\bibitem{Neumann}  Neumann, C. \textit{Die Prinzipien der
Galilei--Newton'schen Theorie} (Leip\-zig: Teubner, 1870).

\bibitem{Grunbaum} Gr\"unbaum, A. \textit{Philosophical Problems of Space
and Time} (New York: Knopf, 1963). 2nd enlarged edition 1973. Dordrecht:
Reidel.

\bibitem{Poinc1898} Poincar\'e, H. \textit{Revue de m\'etaphysique et de
morale}, \textbf{6}, 1, (1898).

\bibitem{Poinc1902}  Poincar\'e, H. \textit{Science et hypoth{\`e}se}
(Paris: Flammarion, 1902).

\bibitem{Poinc1904}  Poincar\'e, H. \textit{Bull. Sci. Math.} \textbf{28},
302, (1904).

\bibitem{Einstein1} Einstein, A. \textit{Ann. Phys.} \textbf{17}, 891 (1905).

\bibitem{Larmor} Larmor, J. \textit{Ether and Matter} (Cambridge: Cambridge
University Press, 1900).

\bibitem{Lorentz} Lorentz, H. A. \textit{Proc. Amsterdam Acad. Sci.} \textbf{%
6} 809 (1904).

\bibitem{Poinc1905} Poincar\'e, H. \textit{Comptes Rendus} \textbf{140} 1504
(1905).

\bibitem{Poinc1906}  Poincar\'e, H. \textit{Rendiconti del Circolo
Matematico di Palermo} \textbf{21} 129 (1906).

\bibitem{Minkowski}  Minkowski, H. \textit{Phys. Z.}  \textbf{10} 104 (1909).

\bibitem{Whittaker1953} Whittaker, E. T. \textit{A History of the Theories
of Aether and Electricity, The Modern Theories 1900 -- 1926}  (London:
Nelson, 1953).

\bibitem{Pais}  Pais, A. \textit{``Subtle is the Lord...'' The Science and
the Life of Albert Einstein}  (Oxford: Oxford University Press, 1982).

\bibitem{Frenkel} Frenkel, J. \textit{Z. Phys.} \textbf{37} 243 (1926).

\bibitem{Kosyakov}  Kosyakov, B. \textit{Theor. Math. Phys.} \textbf{119}
493 (1999).

\bibitem{Gal'tsov} Gal'tsov, D. \textit{Phys. Rev.} \textbf{D 66} 025016
(2002).

\bibitem{Kazinski}  Kazinski, P., S. Lyakhovich, and A. Sharapov \textit{%
Phys. Rev.} \textbf{D 66} 025017 (2002).

\bibitem{Yaremko} Yaremko, Yu. \textit{J. Phys.} \textbf{A 37} 1079 (2004).

\bibitem{Palmieri} Palmieri, C. and B. Vitale
\textit{Nuovo Cimento} \textbf{A 66} 299 (1970).

\end{thebibliography}
\end{document}